\documentclass[12pt]{article}
\usepackage{amsfonts,amssymb,amsmath,latexsym}
\newcommand{\be}{\begin{equation}}
\newcommand{\ee}{\end{equation}}
\newcommand{\bea}{\begin{eqnarray}}
\newcommand{\eea}{\end{eqnarray}}

\newcommand{\ba}{\begin{array}}
\newcommand{\ea}{\end{array}}

\newcommand{\bt}{\begin{tabular}}
\newcommand{\et}{\end{tabular}}

\newcommand{\sump}{\mathop{{\sum}'}}

\newcommand{\fr}{\frac}
\newcommand{\ci}{\cite}
\newcommand{\cl}{\centerline}
\newcommand{\bs}{\bigskip}

\newcommand{\en}{\eqno}

\newcommand{\bbib}{\begin{thebibliography}}
\newcommand{\ebib}{\end{thebibliography}}

\newcommand{\mbb}{\mathbb}

\newcommand{\bm}{\boldmath}
\newcommand{\mb}{\mbox}
\newcommand{\scrs}{\scriptsize}
\begin{document}
\titlepage
\hspace{5cm} {\it LANDAU INSTITUTE preprint 01/06/97}
\vspace{1cm}

\centerline{\large {\bf BKT PHASE TRANSITIONS}}

\centerline{\large {\bf IN TWO-DIMENSIONAL SYSTEMS }}

\cl{\large {\bf WITH INTERNAL SYMMETRIES}}

\vspace{1cm}

\centerline{{\bf S.A.Bulgadaev}\footnotemark{}}
\footnotetext{Extended version of
talk presented
at International Conference "Renorm-Group-96", July 1996, Dubna, Russia.

E-mail: bulgad@itp.ac.ru}

\vspace{0.5cm}

\centerline{L.D.Landau Institute for Theoretical Physics}

\centerline{Kosyghin Str.2, Moscow, 117334, RUSSIA}

\bigskip

\centerline{Abstract}

\vspace{0.75cm}

The Berezinsky-Kosterlitz-Thouless (BKT) type phase
transitions in two-dimensional systems with internal abelian continuous
symmetries  are investigated.
The necessary conditions for they can
take place are:
1) conformal invariance of kinetic part of the model
action;
2) vacuum manifold must be degenerated with abelian and discrete homotopy
group $\pi_{1}.$
Then topological excitations have logarithmically
divergent energy and they can be described by
effective field theories generalizing the two-dimensional euclidean
sine-Gordon theory, which is an effective theory of the initial
$XY$-model.
In particular, the effective actions for the two-dimensional
chiral models on maximal abelian tori $T_G$ of simple compact groups
$G$ are found.
Critical properties
of possible effective theories are
determined and it is shown that they are characterized by the
Coxeter numbers $h_G$ of lattices from the series $\mathbb{A,D,E,Z}$
and can be interpreted as those of conformal field theories
with integer central charge
$C=n,$  where $n$ is a rank of the groups $\pi_1$ and $G.$
A possibility of restoration of full symmetry group $G$ in massive phase
is also discussed.

\newpage

\cl{\large {\bf 1. Introduction}}

\bs

A discovery of  possibility of the phase transition (PT) in
two-dimensional $XY$-model \cite{1} from very beginning has
attracted a great attention of theoreticians due to its
unusual properties. First of all it seems that such PT contradicts
to the well-known theorems by Peierls - Landau \cite{2,3} and
Bogolyubov - Goldstone \cite{4,5} telling us that spontaneous
magnetization and spontaneous  breaking of {\it continuous} symmetry
cannot exist in low-dimensional $(d\le 2)$ systems \cite{6,7}. Secondly,
due to the absence of spontaneous magnetization,
correlation functions in low-temperature phase must fall off algebraically
\cite{8,9}, what means that the whole low-temperature phase has to
be massless.

All these controversies  were brilliantly resolved
in series of papers by Berezinsky \cite{10}, Popov \ci{11}
and by Kosterlitz and Thouless
\cite{12,13}, who have proven for the first time an important
role of topological
excitations - vortices -
with logarithmically divergent energies in this PT.
An existence of vortices is connected with the fact that
the manifold of values of $XY$-model
${\cal M}=S^1$ has nontrivial topology with the discrete abelian
homotopy group
$\pi_1({\cal M})=\mathbb{Z},$  while a logarithmic divergence of the energy
is connected with conformal symmetry of the model.
An account of vortices transforms
continuous compact symmetry $U(1)$ into dual discrete symmetry
$Z_2 \times \mbb Z,$ where $Z_2$ is an automorphism group of $S^1$
corresponding to the reflection symmetry of
$U(1)=S^1.$
Analogous PT take place in other systems with the same symmetry:
two-dimensional SOS and 6-vertex lattice models,  $XXZ$ quantum
spin chains \ci{14} and euclidean sine-Gordon (SG)
model with noncompact field [15-18]. It appears that all these
systems belong to the same critical universality class.
The sine-Gordon model can be considered as an effective
theory of the BKT PT like the Ginzburg - Landau - Wilson theories are
the effective theories
of the PT of II order (see, for example \ci{19}).

The BKT PT can be also connected with some conformal theory, but now there is
a peculiarity. In contrast with usual situation of two-dimensional II order
phase transitions, when an infinite-dimensional conformal symmetry with a
rational central charge $C$ takes
place only at phase transition point \ci{20}, in systems with BKT PT
an infinite-dimensional conformal symmetry with integer central charge
$C=1$ exists not only at PT point (with
logarithmic corrections), but
in a whole low-temperature phase.
Thus we see that the BKT PT is intimately related with two fundamental
properties of the systems: 1) a nontrivial topology described by discrete
abelian  homotopic group $\pi_1$, 2) a conformal symmetry.

It is interesting to consider the BKT PT properties of such systems with
internal symmetries. These systems are related with tori, which are the
natural generalization of circle $S^1$ with necessary properties.
It is clear that the same properties  take place
in two-dimensional chiral models on tori $T^n$ with
$\pi_1(T^n)= {\mbb Z}^n$. This case
effectively reduces to the previous one, since only excitations
with minimal topological charges $e_i = \pm1, i=1,...,n$
are important and such charges with different $i$ do not interact
between themselves. Similar properties have $\sigma$-models on general
tori, connected with arbitrary  nondegenerate lattices $L$ \cite{21}.
But, as has been shown in \cite{21}, besides $T^n,$
there are the maximal abelian tori $T_G$ of the simple compact Lie groups $G$,
which have (in case of simply connected $G$)
$\pi_1({T_G}) = L_v \ne {\mbb Z}^n$ (here $L_v$ is a dual
root lattice of the corresponding Lie algebra ${\cal G}$)
and where excitations with different vector topological charges interact
with each other.

The next question arises naturally:
how the properties of  the above-mentioned topological phase transition
will depend on $G$? This question is important, for example,
for string theory,
where a compactification on $T_G$ (more rigorously, on the simplified tori
$T^n = T_{U(n)}$ or $T_L = {\mbb R}^n/L,$ where $L$ is some nondegenerate
lattice of rank $n$) is considered in different aspects
\ci{22,23}, or for chiral models on $G$ with reduced (or partially broken)
symmetry $G \searrow T_G$ \ci{24}.

In this paper it will be shown that:

1) all  critical properties of
nonlinear $\sigma$-models on compact $T_G$ can be described
in terms of effective field theories with discrete symmetries, generalizing
SG theory;

2) they depend only on
the Coxeter number $h_L$ of the corresponding lattice of topological charges
$L^t;$

3) different classes of universality of the BKT PT are defined by
$\mbb {A,D,E,Z}$ series of the integer-valued lattices;

4) all critical and low-temperature properties of these $\sigma$-models
(except logarithmic corrections at criticality)
can be described by corresponding conformal
theories with integer
central charge $C=n,$ where $n$ is a rank of groups $\pi_1(T_G)$ and $G.$

A possible restoration of the full symmetry group $G$ in massive phase will be
also discussed.

\newpage
\centerline{\large {\bf 2. Nonlinear $\sigma$-models on $T_G$ and vortices.}}

\bigskip

Now we pass to consideration of the
euclidean
two-dimensional chiral field theories on $T_G$,
generalizing
nonlinear $\sigma$-model on a circle $S^1$ or
continuous $XY$-model. Their action has the following
form
$$
{\cal S} = \frac{1}{2\alpha} \int d^2x Tr_{\tau}({\bf t}^{-1}_{\nu}
{\bf t}_{\nu}) =
\frac{(2\pi)^2}{2\alpha} \int d^2x Tr_{\tau}({\bf H}
{\mb {\bm $\phi$}}_{\nu})^2 =
$$
$$
\frac{(2\pi)^2}{2\alpha}N_{\tau}\int d^2x ({\mb {\bm $\phi$}}_{\nu})^2,
\eqno (1)
$$
where
${\bf t}=e^{2\pi i({\bf H} \mb {\scrs \bm $\phi$})} \in T_G,$
${\bf H}= (H_1,...,H_n) \in {\cal C},$ which is the maximal
Cartan subalgebra of the corresponding Lie algebra
${\cal G},$ $[H_i,H_j]=0, \, n$ is
a rank of $G,$ ${\mb {\bm $\phi$}}_{\nu} = \partial_{\nu} \mb {\bm $\phi$},
\, \nu = 1,2,$
and an isotropness  property
of the weight system $\{{\bf w}\}_{\tau}$ of
any $\tau(G)$-representation, which is a consequence of invariance of the
weight systems under discrete Weyl group $W_G,$
is used
$$
\sum_{a} w^a_i w^a_k =
N_{\tau} \delta_{ik}, \; a = 1,..., dim \tau(G).
\en(1a)
$$
It will be convenient below to include a factor $N_{\tau}$ as a normalization
factor into trace $Tr_{\tau}.$ This gives us a canonical euclidean metric
in space of topological charges.

Theories (1), like other two-dimensional chiral models, are invariant
under direct product of right (R) and left (L)
groups $N_G^{R(L)},$ which are a semi-direct product of $T_G$ and $W_G$
$$
N_G = T_G \times W_G.
\en(2)
$$
The group $N_G$ is called a normalizator of $T_G$ and is a symmetry group
of torus $T_G.$

These theories are the multicomponent generalization of $XY$-model, having
properties analogouos to those of $XY$-model:

1) a zero beta-functions $\beta(\alpha)$ due to flatness of $T_G;$

2) non-trivial homotopy group $\pi_1$ and corresponding vortex
solutions.

The classical equations of motion
$$
({\partial}_{\nu})^2 ({\bf H} \mb {\bm $\phi$}) = 0,
\eqno (3)
$$
have vortex solutions in a region $R>r>a,$
where $R$ is a radius of a system and $a$ is a short-wave cut-off parameter
(for example, size  of a vortex core).
$$
{\bf t}(\vartheta) = e^{2\pi i({\bf H} \mb {\scrs \bm $\phi$}(\vartheta))},
\quad
\mb {\bm $\phi$}=\fr{1}{2\pi} {\bf q} (\vartheta)
\eqno (4)
$$
where $\vartheta$ is an angular and $r$ is a radial coordinates
on a plane ${\mbb R}^2,$
${\bf q}{\in}{L_{\tau}^t} = {L_{\tau}}^{-1}$, ${\bf q}$ is
a vector topological charge of a vortex, and $L_{\tau}^t$ is a
lattice of all possible
topological charges of $\tau$ - representation,
${L_{\tau}}^{-1}$ is a
lattice of vectors, inverse to all weights of $\tau$-representation
$$
{\bf q} \in L_{\tau}^t, \quad {\bf w}_a \in \{{\bf w}_{\tau}\},
\quad ({\bf q}{\bf w}_a) \in {\mbb Z}.
\en(5)
$$
For minimal fundamental representations of the simply connected groups
$\tau(G) = min$ a lattice $L_{min}^t = L_v$ and for
adjoint representations
$\tau = ad$ a lattice $L^t_{ad} = {L_r}^{-1} = L_{w^*},$ where
$L_r$ is a root lattice of group $G$ and $L_{w^*}$ is
a  lattice of dual weights or a  weight lattice of dual group $G^*.$
Just these solutions for groups G, such,
that $L^t_{\tau} \supseteq {L_{w}}$, can give
the topological interpretation of
all their quantum numbers \cite{21}.
The energy of these vortices is
logarithmically divergent
$$
E = \fr{(2\pi)^2}{2\alpha}\int (\partial_{\mu} \mb {\bm $\phi$})^2 d^2 x
= \fr{2\pi}{2\alpha}  ({\bf q})^2\ln (R/a).
\en(6)
$$
Due to (2), which defines an effective metric in space of topological charges
\ci{21}, there is a logarithmic interaction between
vortices with different vector topological charges
$$
E = ({\bf q}_1 {\bf q}_2)\fr{2\pi}{2\alpha}
\ln \fr{|{\bf x}_1 - {\bf x}_2|}{a}.
\en(7)
$$
The general $N$-vortex solutions have the next form \cite{20}
$$
\mbox{\bm $\phi$}({\bf x}) = \sum_{i=1}^N {\bf q}_i\fr{1}{\pi}
\arctan(\fr{y-y_i}{x-x_i}),
\en(8)
$$
$$
{\bf q}_i \in L^t_{\tau}, \quad ({\bf q}_i {\bf w}_a)\in {\mbb Z},
\quad (x,y)\in {\mbb R}^2.
$$
The energy of $N$-vortex solution, $E_N,$ with a whole topological
charge $\sum _{i=1}^N {\bf q}_i = 0$ is
$$
E_N= \sum_i E^0_{q_i} + E_{Nint}, \quad
E^0_{q_i}= \fr{1}{2\alpha}C(a)({\bf q}_i{\bf q}_i), \quad
$$
$$
E_{Nint}=
\fr{2\pi}{2\alpha}\sum_{i\ne k}^N ({\bf q}_i{\bf q}_k)
\ln \fr{|x_i-x_k|}{a},
\en(9)
$$
where $E^0_{q_i}$ is a  "self-energy" (or an energy of the core) of
vortex with topological charge ${\bf q}_i$
and $C(a)$ is some nonuniversal constant, depending on type of the core
regularization. Only such solutions give
finite contribution to partition function of the theory ${\cal Z}$.
Since $E_q \sim q^2$
and ${\bf q} \in L^t,$
the maximal
contribution in each $N$-vortex sector  of solutions
will give vortices with minimal $|q|_i$. Thus, in quasi-classical
approximation (or in low T expansion) one can represent a partition
function of theory (5)
$$
{\cal Z}= \int D \mb {\bm $\phi$} \exp(-{\cal S}[\mb {\bm $\phi$}])
\en(10)
$$
in the form of grand
partition function of classical neutral Coulomb gas of vortex solutions with
minimal isovectorial topological charges
${\bf q}_i \in \{{\bf q}\}_{\tau},$ where
$\{{\bf q}\}_{\tau}$ is a set of minimal vectors of lattice $L^t_{\tau}$
$$
{\cal Z}= {\cal Z}_0 {\cal Z}_{CG},\quad
{\cal Z}_{CG}= \sum_{N=0}^{\infty} \frac{\mu^{2N}}{N!} \sump_{\{{\bf q}\}}
{\cal Z}_N(\{{\bf q}\}|\beta).
\en(11)
$$
Here $\sump$ goes over all neutral  configurations of minimal charges
${\bf q}_i \in \{{\bf q}\}_{\tau}$
with $\sum_1^N {\bf q}_i = 0,$
${\cal Z}_0$ is a partition function of free massless isovectorial boson field,
which corresponds to "spin waves" of $XY$-model
$$
{\cal Z}_0 = \int D \mb {\bm $\phi$}\exp (-{\cal S}_0[\mb {\bm $\phi$}]),
\en(12)
$$
$$
{\cal Z}_N(\{{\bf q}\}|\beta)=
\prod_{i=1}^{N} \int d^2x_i \exp(-\beta H_N(\{{\bf q}\}))
\en(13)
$$
$$
H_N(\{{\bf q}\})= \sum_{i<j}^{N} ({\bf q}_i{\bf q}_j)D(x_i-x_j),
\en(14)
$$
$$
D(x)= \int \frac{d^2k}{(2\pi)^2}(e^{i({\bf k}{\bf x})}-1)f(ka)/k^2
\mathrel{\mathop {\sim}\limits_{|x|\gg a}} \frac{1}{2\pi} \ln|x/a|
\en(15)
$$
where
$$
\mu^2 = a^{-2} y^2 det,\quad y^2 = e^{-E^0_q}
\en(15a)
$$
is a chemical activity of Coulomb gas,
$det$ is a determinant of the fluctuations
over one vortex solution (futher we will suppose that it is equal to some
constant of order O(1) and assume that $det =1$),
$$
\beta = 2(\pi)^2/\alpha,
\en(15b)
$$
$f(ka)$ is a regularisator such that
$$
lim_{k \to 0} f(ka)=1, \quad lim_{k \to \infty} f(ka) =0.
$$
How the account of vortices reduces the initial symmetry group $N_G$ of the
$\sigma$-model becomes obvious in the next section.

\bs

\centerline {\large \bf 3. Duality of the compact and noncompact theories.}

\bigskip

In case of $XY$-model in long-wave quasi-classical approximation
there is an  important connection between partition function of compact
chiral theory (1)
on $S^1$ and a partition function of the
noncompact SG theory [15-17] with action  (modulo ${\cal Z}_0$)
$$
{\cal S}_{SG}= \int d^2x \left(\fr{1}{2\beta}(\partial_{\mu} \phi)^2 -
{\mu}^2 Cos(\phi)\right).
\en(16)
$$
This action has explicit invariance under dual discrete group
$Z_2\times {\mbb Z}.$
Analogous connection exist between compact chiral models on $T_G$ and
noncompact generalized SG theories.

To see this we note that
the grand partition function ${\cal Z}_{CG}$ from (11) is in its turn
equivalent to partition function of noncompact
scalar isovectorial field theory
$$
{\cal Z}_{CG} = \int D\mb {\bm $\phi$} e^{-{\cal S}_{eff}}, \quad
{\cal S}_{eff} = \int \fr{1}{2\beta}(\partial \mb {\bm $\phi$})^2
-\mu^2 V(\mb {\bm $\phi$}),
\en(17)
$$
$$
V(\mb {\bm $\phi$}) =
\sum_{\{{\bf q}\}} \exp i({\bf q} \mb {\bm $\phi$}).
\en(18)
$$
where $\sum_{\{{\bf q}\}}$ goes over the set of minimal topological
charges, and $\mb {\bm $\phi$} \in {\mbb R}^n$ \cite{25}.
Strictly speaking the theories (17) with arbitrary parameters $\mu$
and $\beta$ are more general than initial $\sigma$-models (1). The last
have only one parameter, a coupling constant $\alpha.$ The representation
of $\sigma$-models
in the form (11,17) gives their embeddeding into general theories (17),
since there are the restrictions (15a,b),
relating parameters $\mu$ and $\beta.$
This fact will be important later
when a possible increasing  of the symmetry of $\sigma$-models will be
discussed (section 6).

Since the set of minimal charges $\{ {\bf q}\}_{\tau}$ is invariant under
dual Weyl group $W_{G^*}$, we see that the account of vortices reduces
the initial symmetry group $N_G$ into discrete dual group
$W_{G^*} \times L_q^{-1}.$ Here a lattice $L_q^{-1}$ is
a periodicity lattice   of potential $V$ and, consequently,
is inverse to all ${\bf q} \in \{{\bf q}\}.$ It follows from their
definitions that $L_q^{-1} =  L_{\tau}.$ This dual group generalizes
the dual group $Z_2 \times {\mbb Z}$  of $XY$-model.

Thus, in this semiclassical and long wavelength approximation
{\it compact}
theory on a torus $T_G$ with continuous symmetry $N_G$ appears to be
equivalent (again modulo ${\cal Z}_0$)
to {\it noncompact}
theory with periodic potential and an infinite discrete symmetry.
These potentials contain the sum over all minimal vectors $\{{\bf q}\}$
and can coincide with characters of some representations of group $G.$
For example, in case
of $L^{-1}_{\tau} = L_v$ the sum in (18) goes over all dual minimal
roots.  Thus the
corresponding potentials $V$  for simply laced groups  from series $A,D,E$
coincide
with characters of adjoint representations of these groups (modulo some
constant, corresponding to zero weight). In this case the general
theories (17) can describe systems  with symmetry $G$ broken to
$N_G$ \ci{25}.

The noncompact theories (17) can be considered also as corresponding
linear $\sigma$-models.
Thus we have shown that {\it compact nonlinear} $\sigma$-models on $T_G$ are
equivalent in this approximation to {\it noncompact linear} $\sigma$-models
on Cartan tori of dual group $T_{G^*}.$

For further consideration we need to classify all possible effective theories
of this type. It follows from (17,18) that they are determined by the set of
minimal vectors $\{{\bf q}\}$ of lattice $L^t_{\tau},$ which satisfies
the next restriction
$$
L_{w^*} \supseteq L^t_{\tau} \supseteq L_v.
\en(19)
$$
For $\tau = min$ a lattice
$L^t_{\tau} = L_v,$  for $\tau =ad$ a lattice $L^t_{\tau} = L_{w^*}.$
The lattices $L_v$ and $L_{w^*}$ differ by a factor, which is isomorphic
to the centre $Z_G$ of group $G$
$$L_{w^*} / L_v = Z_G.$$  Thus the set $\{{\bf q}\}$ can vary from
the set of minimal vectors (it forms the so called Voronoi region or
Wigner - Seitz cell of the corresponding lattice) of the weight lattice
till that of the root lattice.  All possible cases are determined by
subgroups of the centre $Z_G.$ For groups $G$ with $Z_G=1$ the lattices
$L_v$ and $L_{w^*}$ coincide.

\bigskip

\cl{\large \bf 4. Phase transition in chiral models on $T_G$.}

\bigskip

In this section we consider topological phase transitions in chiral models
on $T_G$ using  above obtained approximate equivalence of these compact
theories with noncompact generalized SG field theories (17).
These field theories can be considered as
effective theories for topological phase transitions in chiral models
defined on $T_G$ like the sine-Gordon theory for $XY$-model [15-17] and
the Ginzburg-Landau-Wilson theories for
the II order phase transitions \ci{19}.

The investigation of the BKT type phase
transition for all effective field theories of the form (17)
was done by the renorm-group method in \cite{25}. It was shown there
that the new critical properties can have only theories connected with
even integer-valued lattices of $\mbb {A,D,E}$ type. They have a structure
of root lattices of the corresponding simple groups $G$ from simply laced
series $A,D,E.$ All theories connected with other lattices have the same
critical properties as SG theory or its superpositions  connected with
lattice ${\mbb Z}^n.$
Here we give brief description and write out obtained  results,
paying the main attention to the symmetry and universality properties.

Under renormalization transformations both parameters $\mu$ and
$\beta$ are renormalized. It is convenient to introduce two
dimensionless parameters
$$
(\mu a)^2=g, \, \delta = \frac{\beta q^2 - 8\pi}{8\pi}
\en(20)
$$
where $q^2$ is a square of the norm of the minimal vector topological
charges from $\{{\bf q}\}.$ The theories (17) are renormalizable
only if the vectors from  $\{{\bf q}\}$ belong to some lattice (here
$L^t_{\tau}$). A new critical properties can appear only if a geometry
of $\{{\bf q}\}$ is such that each vector ${\bf q} \in \{{\bf q}\}$
can be represented as a sum of two other vectors from $\{{\bf q}\}$ \ci{25}.
The last property is very restrictive and concides with a definition
of the root systems $\{{\bf r}\}$ of simple groups from series
$A,D,E$ \ci{25}  or
with a definition of the root set of the even integer-valued
(in some scale) lattices of $\mbb {A,D,E}$ types \ci{26}. The sets of
minimal roots (or minimal dual roots) of all simple groups belong to
four series of integer-valued lattices $\mbb {A,D,E,Z}.$ For theories with
sets $\{{\bf q}\} \notin \mbb {A,D,E,}$ all critical properties will be
the same as for theories with $\{{\bf q}\} \in {\mbb Z}^n$ \ci{25}.

The RG equations for theories (17) with $\{{\bf q}\}$ from these lattices
have the next form \ci{25} (here $G=A,D,E$)
$$
\fr{dg}{dl}=-2\delta g + B_{G}g^2, \quad
\fr{d\delta}{dl}= - C_{G}g^2.
\en(21)
$$
Here $B_{G}=\pi\theta_{G},$
$\theta_{G}$ is the multiplicity of the reproduction of $V(\mb {\bm $\phi$})$
under renormalization of (17) or the number of times
by which each root can be represented as a sum of two other roots,
and $C_{G}=2\pi K_G$, where $K_G$ is the value of the second Casimir
operator in adjoint representation (where ${\bf w}_a = {\bf r}_a$)
$$
\sum_{a} r^a_i r^a_j= K_G \delta_{ij}.
\en(22)
$$
The RG equations of type (21) with coefficients corresponding to the case
$G= A_2$ have been obtained firstly in \ci{27} under investigation of the
melting of the two-dimensional triangle lattice.

The eigenvalue of the second Casimir
operator $K_G$  for groups from
series $A,D,E$ can be expressed in terms of the corresponding Coxeter
number $h_G$
$$
K_G=2h_G, \quad h_G=
\frac{\mbox{(number of all roots)}}{\mbox{(rank of group)}}.
\en(23)
$$
This definition of the Coxeter number coincides with that of the Coxeter
number of the corresponding lattices from series $\mbb {A,D,E}.$
The coefficient $B_G$ can be calculated by different methods, and
can be expressed also through the Coxeter number
$$
\theta_G = K_G - 4 = 2(h_G-2)
\en(23)
$$
Thus we see that all coefficients of RG equations depend only on
the Coxeter number $h_G$ or on the second Casimir value $K_G$.
The RG equations (21) have two separatrices \ci{25,27}:
$$
u_{1,2}\equiv (g/\delta)_{1,2} = \fr{1}{2C_G}
\left[\pm (B_G^2 + 8 C_G)^{1/2} - B_G \right],
\en(24)
$$
where $u_1$ corresponds to the phase separation line.
The critical exponent $\nu_G,$ determining divergence of the correlation
length $\xi$
$$
\xi \sim a\exp(A\tau^{-\nu_G}), \quad \tau = \fr{T-T_c}{T_c},
$$
is given by the next expression
$$
\nu_G= 1/\kappa_G  = u_1\left[(B_G/C_G)^2+8/C_G)\right]^{-1/2},
\en(25)
$$
where $1/\kappa_G$ is the Lyapunov index of the separatrix 1 \ci{25}.
Substituting corresponding values one finds
the declinations of two separatrices
$$
u_{1,2}=
\left\{
\begin{array}{c}
1/\pi h_G,\\
-1/2\pi.\\
\end{array}\right.
\en(26)
$$
Note that $u_{2}= -1/2\pi$ does not depend on $G$ and is universal constant.
This fact is very important for a possibility of restoration
of full symmetry group $G$ (see below section 6).
A schematic phase diagram is depicted on Fig.1.

\begin{picture}(400,150)(-80,0)
\put(100,0){\vector(1,0){100}}
\put(100,0){\vector(0,1){100}}
\put(100,0){\line(-1,0){100}}
\put(100,0){\line(-1,2){50}}
\put(100,0){\line(2,1){100}}
\put(200,20){I}
\put(50,20){III}
\put(120,70){II}
\put(210,0){$\delta$}
\put(100,110){g}
\put(97,-15){0}
\put(60,100){2}
\put(170,40){1}
\qbezier[100](70,80)(120,30)(195,25)
\end{picture}

\bs

\cl{ Fig.1. A schematic phase diagram}

\bs

The dashed line of the initial values  corresponds to the initial
$\sigma$-model. This line is defined by the dependencies of the parameters
$\beta$ and $\mu$ on coupling constant $\alpha$ (equations (15a,b)).
To low-temperature phase (decompactified, massless) corresponds
region I, other
regions answer to a high-temperature (compact, massive) phase.
In the region I the
correlation length $\xi = \infty$, and in the region II, near separatrix 1,
$$
\xi \sim ae^{A\tau^{-\nu_G}}, \;
\nu_G =2/(2+h_G) = 4/(4+K_G.)
\eqno (27)
$$
Using known values for the Coxeter  numbers $h_G$ and a geometry of the
minimal dual root sets,
we obtain the following expressions for critical exponents ${\nu}_G$
(see Table 1) \cite {25}

\centerline{Table 1}

$$
\ba{|c|c|c|c|c|c|c|c|c|c|}
\hline
{} & {} & {} & {} & {} & {} & {} & {} & {} & {}\cr
G & A_n & B_n & C_n & D_n & G_2 & F_4 & E_6 & E_7 & E_8\cr
{} & {} & {} & {} & {} & {} & {} & {} & {} & {}\cr
\hline
{} & {} & {} & {} & {} & {} & {} & {} & {} & {} \cr
\nu_G & \fr{2}{n+3} & \fr{1}{n} & \fr{1}{2} & \fr{1}{n} & \fr{2}{5}
& \fr{1}{4} & \fr{1}{7} & \fr{1}{10} & \fr{1}{16} \cr
{} & {} & {} & {} & {} & {} & {} & {} & {} & {} \cr
\hline
\ea
$$

\bs

It is interesting to note, that groups $D_{16} = O(32)$ and $E_8$, used in a
construction of the
anomaly-free theories of strings \cite{22}, have identical ${\nu}_G$
(together with $A_{29}$).
  The greatest number of possible values ${\nu}_G$ is given by
$A_n$: $1/k$ and $2/(2k + 1),$ where $k$ is an integer.
For the theories with $V$,  containing a set of the minimal roots,
all indices remain the same, except ${\nu}_{B_n}$ and
${\nu}_{C_n}$
which interchange themselves due to  mutual duality of their groups.

\bs

\cl{\large{\bf 5. Low temperature phase and conformal symmetry}}

\bs

The equality $ \xi \to \infty $ everywhere in low-temperature phase
means an existence  of conformal symmetry in it at large distancies.
It can be seen also from renormalized effective action
${\cal S}_{eff}$ of the theory, which has in IR limit the next form
$$
{\cal S}_{eff} = \int d^2x \frac{1}{2 {\bar \beta}}
(\partial \mb {\bm $\phi$})^2,
\en(28)
$$
where $\bar \beta$ is the IR limit value of the renormalized $\beta(l)$
$$
\bar \beta = lim_{l\to \infty} \beta(l)
\en(29)
$$
At the PT point $\bar \beta = \beta^* = 8\pi/q^2_{min}.$
In other points of low T
phase $\bar \beta$ depends on initial values of the system.
It is well known that action (28) describes free conformal theory
with central charge $C = n,$ here $n$ is a rank of group $G.$
It means that long-wave low T properties of $\sigma$-models,
defined on different
torus $T_G,$ are the same for all groups with equal rank $n.$
Only logarithmic corrections at PT pont will depend on group $G$ through
the Coxeter number $h_G.$
In this relation it becomes clear why
the critical indexes depend only on $h_G$ or $C_2^{adG}.$  It agrees
with the fact that on compact
groups all quantum conformal anomalies  depend also only on $h_G$ (or
dual $\tilde h_G$) \ci{28}. In this connection it is interesting, that
$\nu_G$ coincides with a "screening" factor in the formula for
the central charge $C_k$ of affine algebra $\hat G$ \cite{28}
$$
C_k = \frac{k}{k+h_G} dimG
\eqno (30)
$$
at a level $k = 2$ or in the formula for $C_k$ in "coset" realization
${{\hat G}_k \otimes {\hat G}_1}/{\hat G}_{k+1}$ of
minimal unitary conformal models \cite {29}
$$
C_k = n\left(1-\frac{h_G(h_G+1)}{(h_G+k)(h_G+k+1)}\right)
\eqno (31)
$$
at a level $k=1$.

The fact that low T phase is effectively free field phase permits the
calculation of correlation functions. For example, for correlation
functions of the exponentials  one obtains the next expression
$$
\left< \prod_{s=1}^{t} exp(i({\bf r}_s \mb {\bm $\phi$}(x_s)))\right> =
\prod_{i\ne j}^{t} \left|\fr{x_i-x_j}{a}\right|^
{\bar \beta ({\bf r}_i{\bf r}_j)/2\pi},
\quad \sum_{i=1}^{t} {\bf r} = 0.
\en(32)
$$
At the PT point (where  $\bar \beta = \beta^{*} = 8\pi/q^2 = 4\pi$) an
additional logariphmic factor, related with the "null charge" behaviour
of $g$ and $\delta$ on the critical separatrix, the phase separation line,
appears in them:
$$
\prod_{i\ne j}^{t} \left(\ln \left|\fr{x_i-x_j}{a}\right|
\right)^{\beta^{*}({\bf r}_i{\bf r}_j)/2\pi A_G}=
\prod_{i\ne j}^{t} \left(\ln \left|\fr{x_i-x_j}{a}\right|
\right)^{h_G \cos({\bf r}_i{\bf r}_j)},
\en(33)
$$
where $A_G= 4/h_G$ is a coefficient in RG equations for $\delta$ on
the critical separatrix.

\bs
\cl{\large{\bf 6. Massive phase, asymptotic freedom }}
\cl{\large{\bf and global symmetry}}

\bs

The regions II and III answer in IR-limit to  the high-temperature
(in statistical mechanics language), massive (in field theory language),
phase.
In UV-limit the region III will be asymptotically free. A separatrix
2 with decline  $u_2 = -1/2\pi$ plays also an important role.
In UV-limit it is a boundary of the asymptotically free region III.
There is also another possibility of the enhancement of the symmetry of
the initial nonlinear  $\sigma$-model
on this separatrix.  $\sigma$-model (1) has at classical level
two symmetries:
1) scale (or conformal) symmetry, 2) isotopic global symmetry
$N_G = T_G \times W_G.$
At quantum level the first symmetry is, in general,
spontaneously broken in IR region
by vortices (see (11)). For this reason $\sigma$-model has in massive phase
a finite correlation length $\xi \sim  m^{-1}$, where  $m$ is a
characteristic mass scale of the theory. This mass must depend on the coupling
constant $\alpha$ or $\beta.$  The behaviour of $m$ near PT point is described
by formula (27), where
$$
\tau \sim \fr{\alpha - \alpha_c}{\alpha_c}.
$$
There is another region in massive phase, the separatrix 2, where
$m(\alpha)$ can be found.
Since in IR-limit this separatrix attracts all trajectories in massive
(or high-temperature) phase, it is very important to know an effective
mass scale on it.
In main approximation on g  it is given by pole in RG equation  \cite{30}
or by the formula
$$
m \sim \Lambda \exp(-\int^g dx/ \beta(x))
$$
where $\Lambda \sim a^{-1}$ is UV cut-off parameter in momentum space,
$\beta(x))$ is a $\beta$-function on the separatrix 2
$$
\beta(g) = 2\pi g^2 K_2^G/2 = 2\pi g^2{h_G}.
$$
Therefore one obtaines
$$
m \sim\ a^{-1} \exp{(-1/2\pi gh_G)}.
$$
The numerical factor in $\beta$-function can vary,
in dependence on a normalization of the coupling constants,
but the fact that on a separatrix 2 $\beta \sim K_2^G \sim h_G$ is
a corollary of the property
that the declination of this separatrix $u_2$ {\em does not depend on} $G.$

This expression for a mass scale on separatrix 2,
depending only on $K_G,$ coincides with those for chiral models on groups
\cite {30} and with those, obtained
from an exact solution of the appropriate fermion theories in main
approximation on $g$ \cite{31}
$$
m \sim \Lambda \exp (-2\pi/(gK_G/2)),
$$
Thus, it appears, that a mass scale on separatrix 2 coincides
(at least for $G=A,D,E$)
with that for $G$-invariant theories
(chiral and fermionic), connected with simple Lie groups $G,$
and can be expressed only through the Casimir
operator $K_G$
by the universal formula (ours $g \to g/(2\pi)^2$)
$$
m \sim \Lambda e^{-4\pi/K_G g}
$$
It means that on this separatrix  the theory (17) can become $G$-invariant.
This can
be seen also from the equivalence of the effective field theories (17)
in  cases $G=A_{n-1} = SU(n), G = D_n = O(2n), G = E_{6,7}$ to the
fermion theories with the same glodal symmetry groups $G$ [25a].

From here it follows, that in massive phase of chiral theory
on $T_G$ ($G=A,D,E$) in minimal representation (when $L^t_{min} = L_v$)
there is strong dependence
of the mass scale on coupling constant,
which interpolates between formula (27) near the PT point and formula (34)
near the separatrix 2.
The first region corresponds to symmetry
of $T_G,$ which is a torus normalisator $N_G= T_G \times W_G,$ while
the second one corresponds to more symmetric,  $G$-invariant,
situation.
Analogous crossover in $m(\alpha)$ takes place in $\sigma$-models on other
groups and in other representations, but the symmetry properties in two
limiting regions remain not so clear.

This work was supported by RFBR grants 96-02-17331 and 96-15-96861.

\begin{thebibliography}{99}

\bibitem{1}  Stanley H.E., Kaplan T.A., Phys.Rev.Lett.{\bf 17} (1966) 913.
\bibitem{2} Landau L.D., ZETP {\bf 7} (1937) 627.
\bibitem{3}  Peierls R.E., Ann.Inst.Henry Poincare {\bf 5} (1935) 177.
\bibitem{4} Bogolyubov N.N., Selected works, v.{\bf 3}, Kiev,
Naukova dumka, 1971.
\bibitem{5} Goldstone J., Nuovo Cim. {\bf 19} (1961) 154.
\bibitem{6}  Mermin N., Wagner H., Phys.Rev.Lett., {\bf 17} (1966) 1133.
\bibitem{7}  Hohenberg P.C., Phys.Rev. {\bf 158} (1967) 383.
\bibitem{8} Rice T.M., Phys.Rev.{\bf 140} (1965) 1889.
\bibitem{9}  Jancovici B., Phys.Rev.Lett. {\bf 19} (1967) 20.
\bibitem{10}  Berezinsky V.L., ZETP {\bf 59} (1970) 907,
{\bf 61} (1971) 1144.
\bibitem{11}  Popov V.N., Feynman integrals in quantum field theory
and statistical mechanics, Moscow. Atomizdat. 1976.
\bibitem{12}  Kosterlitz J.M.,Thouless J.P., J.Phys. {\bf C6} (1973) 118.
\bibitem{13}  Kosterlitz J.M., J.Phys. {\bf C7} (1974) 1046.
\bibitem{14} Baxter R.J., Exactly Solved Models in Statistical Mechanics,
Academic Press, 1982.
\bibitem{15}  Jose J.,Kadanoff L., Kirkpatrick S.,Nelson D.,
Phys.Rev. {\bf B16} (1977) 1217.
\bibitem{16} Wiegmann P.B., J.Phys.{\bf C11} (1978) 1583.
\bibitem{17} Ohta T., Prog.Theor.Phys. {\bf 60} (1978) 968.
\bibitem{18} Amit D.J., Goldschmidt Y.Y., Grinstein G.,
J.Phys. {\bf A13} (1980) 585.
\bibitem{19} Patashinskii A.Z., Pokrovskii V.L., Fluctuation theory of phase
transitions, Nauka, Moscow, 1982.
\bibitem{20}   Belavin A.A., Polyakov A.M., Zamolodchikov A.B.,
    Nucl.Phys. {\bf B241} (1984) 333;
Dotsenko Vl.S.,Fateev V.A.,
Nucl.Phys.{\bf B240} (1984) 312, {\bf B251} (1985) 691;
Friedan D., Qiu Z., Shenker,
Phys.Rev.Lett. {\bf 53} (1984) 1575;
Andrews G.E., Baxter R.J., Forrester P.J.,
J.Stat.Phys. {\bf 35} (1984) 193;
Huse D.A., Phys.Rev. {\bf B30} (1984) 3908.
\bibitem{21}
Bulgadaev S.A.,
Pisma v ZETP {\bf 63} (1996) 758 (JETP Letters {\bf 63} (1996) 796);
Bulgadaev S.A., On topological interpretation of
quantum numbers, Landau Institute preprint, 1997; hep-th/9901035,
to be published in JETP N10 (1999).
\bibitem{22}   Green M., Schwarz J.H., Witten E., Superstrings Theory,
Cambridge, 1988, vol.{\bf 1,2};
Kogan Ya.I., Pisma v ZETP, {\bf 45} (1987) 556;
Gross D., Klebanov I., Phys.Rev. {\bf D} (1991),
in Proceedings of the Trieste Spring School "String Theory and
Quantum Gravity'91",  Workshop ICTP, Trieste, Italy,
World Scientific, 1991.
\bibitem{23} Banks T., Fischler W., Shenker S.H., Susskind L.,
Phys.Rev.{\bf D55} (1997) 5112; Banks T., Seiberg N., Nucl.Phys. {\bf B497}
(1997) 41.
\bibitem{24}  Bulgadaev S.A., On decompactification
transition in two-dimensional $\sigma$-models. Talk given at
    International conference
    "Conformal Field Theories and Integrable Models",
    Chernogolovka, Russia, 24-29 June, 1996. Extended version in
    Landau Institute preprint 02/06/97, 1997.
\bibitem{25}
Bulgadaev S.A.,
Phys.Lett. {\bf 86A} (1981) 213;
Teoret.Matem.Fizika {\bf 49} (1981) 7; Nucl.Phys.{\bf B224} (1983) 349;
Pisma v ZETP {\bf 63} (1996) 743 ( JETP Letters {\bf 63} (1996) 780).
a) Bulgadaev S.A., Nucl.Phys.{\bf B224} (1983) 349;
On boson-fermion equivalence for some exceptional groups. Landau Institute
preprint, 1997.
\bibitem{26}  G.H.Conway, N.J.A.Sloane, \em Sphere Packing, Lattices and
Groups, \em vol.I,II. Springer-Verlag, (1988).
\bibitem{27}
Nelson D.R., Phys.Rev. {\bf B18} (1978) 2318;
Nelson D.R., B.I.Halperin, Phys.Rev.{\bf B19} (1979) 2457;
\bibitem{28}Kac V.N., Infinite dimensional Lie algebras, Cambridge
University Press, 1990.
\bibitem{29} Goddard P., Kent A., Olive D., Phys.Lett. {\bf B152} (1985) 88,
Commun.Math.Phys. {\bf 103} (1986) 105.
\bibitem{30}  Polyakov A.M., Gauge Fields and Strings, Harwood Academic
Publishers, 1987.
\bibitem{31} Destri C., de Vega H.J., Preprint CERN-TH, 4895/87 (1987).
\ebib

\end{document}